\documentclass[runningheads]{llncs}
\usepackage[T1]{fontenc}
\usepackage{amsmath,amssymb,subfigure}
\usepackage{algorithm,algorithmic}
\usepackage{graphicx}
\usepackage{comment}
\usepackage{setspace}
\usepackage{longtable}
\usepackage{times,epsfig}
\usepackage{hyperref}
\usepackage{soul}
\usepackage{mathtools}
\usepackage[title]{appendix}

\begin{document}

\title{Centrality Measures : A Tool to Identify Key Actors in Social Networks}

\author{Rishi Ranjan Singh}
\authorrunning{Rishi Ranjan Singh}
\institute{
Department of Electrical Engineering and Computer Science\\
Indian Institute of Technology, Bhilai\\
Chhattisgarh, India. \newline
\email{rishi@iitbhilai.ac.in}}

\maketitle

\begin{abstract}

Experts from several disciplines have been widely using centrality measures for analyzing large as well as complex networks. These measures rank nodes/edges in networks by quantifying a notion of the importance of nodes/edges. Ranking aids in identifying important and crucial actors in networks. In this chapter, we summarize some of the centrality measures that are extensively applied for mining social network data. We also discuss various directions of research related to these measures. 
\end{abstract}

\section{Introduction}

Social networks are an abstraction of real-world social systems where people are represented as nodes and social relationship among them are portrayed as links between nodes. The number of nodes and links in a network are referred as the \textit{order} and \textit{size} of that network. The order of social networks vary a lot. It may be as small as in two digits, for example, 
Zachary's karate club\cite{Zachary:1977}. It may be as large as in millions. Orkut, Flickr, LiveJournal~\cite{Mislove:2007}, Facebook~\cite{Traud:2012}, Twitter, Instagram, etc. are examples of popular online social networks of that order. The number of active Facebook users has been reported in few billions by \textit{https://www.statista.com/} in August 2020. These networks are dynamic in nature and are continuously changing at a fast pace. Every hour, several users are joining or leaving online social network platforms, forming new connections or blocking/deleting older relationships, giving rise to addition/deletion of node and links in the corresponding networks. \\

Social network analysis is a sub area within Network Science and Analysis where researchers attempt mining social network data for various applications. The books by Wasserman et al.~\cite{Wasserman:1994}, Carrington et al.~\cite{Carrington:2005}, Scott and Carrington~\cite{Scott:2011}, and Knoke and Yang~\cite{Knoke:2019} may be referred for basic and detailed understanding of social network analysis. A book written in popular-science style by Freeman~\cite{Freeman:2004} discusses the development of social network analysis area.\\

There are several research problems related to analysis of complex networks which are also studied for social network analysis. For example: identifying important nodes and edges in a given network by defining and applying centrality measures; partitioning networks into densely connected sub-networks which are sparsely connected with each other by detecting community structures; understanding spreading patterns of ideas, memes, and information by studying information diffusion models, guessing which non-adjacent nodes have high probability of becoming adjacent in future by predicting links, etc.\\

This chapter aims to to summarize some of the centrality measures that are extensively applied for mining social network data and identifying key actors. Experts from several disciplines have been widely using centrality measures for analyzing large as well as complex networks. These measures rank nodes/edges in networks by quantifying a notion of the importance of nodes/edges based on a given application. Therefore, the definition of importance is application specific and it changes from one application to another. Ranking aids in identifying important and crucial actors in networks. In the last two decades, several interdisciplinary studies evolved just around the use of these measures to extract information from underlying network data. A major portion of those research works is concerned with selecting the best of the available centrality measures for a particular application. Several other measures have been defined by either generalizing or extending the classical centrality measures. \textit{Group-centrality measures}~\cite{Everett:1999} are a variant of centrality measures where the goal is to rank subsets of nodes by computing collective centrality measures of subsets. \textit{Hybrid-centrality measures} are those measures that are defined by combining different simple centrality measures for better performance.\\

This chapter starts with the basic notion of centrality measures. Then, we cover the definition of the traditional and few other popular centrality measures for social network analysis. We briefly mention algorithms to compute and estimate these measures. Next, we discuss various directions of research related to centrality measures. Afterwards, we summarize a handful applications of various centrality measures for analyzing real-world social networks. Finally, we conclude the chapter with a discussion on some future directions and open problems.

\section{Centrality Measures} 
Centrality measures are network analysis tools to identify most powerful, central or important people /relationship in social networks. In this section, we discuss the traditional centrality measures which are not only popular in social networks but across all types of networks. Further, we also summarize few other centrality measures related to social networks. We use the following notations throughout the chapter. Let $G=(V,E)$ be a social network, where $V$ denotes the set of nodes representing people and $E$ denotes the set of links representing relationships between people. For simplicity, we discuss every centrality measure in this section in the context of undirected and unweighted social networks. It is trivial to extend these for weighted as well as directed social networks. Let $n$ be the number of nodes (order), i.e. $|V|=n$ and $m$ be the number of links (size), i.e. $|E|=m$. Let $A$ be the $n\times n$ adjacency matrix of $G$ where the relationship between node $i$ and node $j$ is denoted by $a_{ij}$, an entry in $A$.

\subsection{Traditional Centrality Measures\label{trad}} In this section, we discuss four traditional centrality measures: degree, closeness, betweenness, and eigenvector.  

\subsubsection*{Degree Centrality:} This centrality measure quantifies direct friendship support available to a node in social networks. As per this notion of power, a node's importance is assumed to be proportional to its degree~\cite{Freeman:1979}. The degree centrality of a node $i$, $DC(i)$ is defined as $$DC(i)=\sum_{j\in V\setminus\{i\}} a_{ij}.$$
where $a_{ij}$ denotes the adjacency relationship between node $i$ and $j$. The normalization factor is $n-1$ i.e., these values can be normalized by dividing the degree of nodes with $n-1$, where $n$ denotes the order of networks.

It is a notion of popularity in social networks. Nodes with a large number of relationships are powerful and central according to this measure and exhibits higher following, strength and emotional support available. Such nodes are also highly exposed to flowing information or spreading disease in networks. Nodes with a small number of degree are not very popular and represent introvert personalities. The limitation of this measure is its local view of the network topology due to which it uses only limited local knowledge to decide the importance.  

\subsubsection*{Closeness Centrality:} This measure has been known as \textit{status} of a node since 1959~\cite{Harary:1959}. Freeman~\cite{Freeman:1979} in 1979 termed it as closeness centrality. According to him, power of a person in a social network in terms of closeness centrality is inversely proportional to the sum of its distance to all the other persons in that social network. The closeness centrality of a node $i$ is computed as
$$CC(i)=\frac{1}{\sum_{j\in V\setminus\{i\}} d_{ij}},$$ where $d_{ij}$ denotes the shortest path length from node $i$ to node $j$. $d_ij$ is also known as geodesic distance from node $i$ to $j$. The normalization factor is $\frac{1}{n-1}$ i.e., this measure can be normalized by multiplying the values with $n-1$. Recall, $n$ denotes the order of social networks. 

Closeness centrality doesn't work in disconnected networks. Therefore, \textit{harmonic centrality}~\cite{Opsahl:2010,Boldi:2014} may be used in its place which is a highly correlated measure with closeness centrality. Harmonic centrality measure assumes importance proportional to the sum of inverse of distances. It is defined as $$HC(i)=\sum\limits_{j\in V\setminus\{u\}} \frac{1}{ d_{ij}}.$$

The closeness centrality of a node quantifies the average distance to all other nodes in the network from that node. This notion is useful to identify those nodes which receive any information originated anywhere in the network in the least expected time. It is due to a smaller expected length from the originating node. Vice-versa, any information originating at high closeness central nodes takes small amount of the expected time to reach to all other nodes. As the information reaches to closeness central nodes quickly, therefore, it is of high fidelity, i.e. with low noise in information. On the negative aspect, these nodes are prone to get infected from a spreading disease in the network faster than other nodes due to expected shorter distance from the seed nodes for diseases and vise-versa. 

\subsubsection*{Betweenness Centrality:}  Bavelas~\cite{Bavelas:1948} defined the notion of importance of a point in communication networks proportional to the number of shortest paths between other points that are passing through that point. It was termed as \textit{point centrality}. Later, Anthonisse~\cite{Anthonisse:1971} and Freeman~\cite{Freeman:1977} introduced independently the definition of betweenness centrality. The betweenness centrality version of power and importance of a node is assumed to be proportional to the fraction of shortest paths between all possible pairs of nodes that are passing through that node~\cite{Freeman:1979}. The betweenness centrality of a node $i$ is, $$BC(i)=\sum\limits_{i,j,k \epsilon V: \{i,j,k\}=3 }\frac {\sigma_{jk}(i)}{\sigma_{jk}},$$ where $\sigma_{jk}(i)$ denotes the number of shortest paths from node $j$ to $k$ which are passing through node $i$ and $\sigma_{jk}$ denotes the total number of shortest paths from node $j$ to $k$. The normalization factor is $\binom{n-1}{2} =\frac{(n-1)(n-2)}{2}$ i.e $$BC(i)=\frac{2}{(n-1)(n-2)} \sum\limits_{i,j,k \epsilon V :  \{i,j,k\}=3 }\frac {\sigma_{jk}(i)}{\sigma_{jk}}.$$

The definition of this measures is based on the assumption that transportation and communication happens through shortest paths between nodes. In social network, this measure represent the brokerage power of a person. It is also a good indicator of the expected amount of communication load a node has to handle. A person with high betweenness centrality has higher control over the information flowing across the network. At the same time, that person is heavily loaded due to the reason that a major fraction of the information flow across the network is happening through him/her. In some types of flow networks (e.g. Power Grid networks, Gas Line networks, and Communication networks) heavy load may also attract frequent demand of maintenance and such nodes are relatively more prone to fail resulting in major breakdown in the network system. Several studies tried to replicate the phenomena of cascading failure\cite{Kinney,Buldyrev,Lin,Motter} and observed that faults at high load nodes may cause a cascading failure and finally breakdown of the whole system.

\subsubsection*{Eigenvector Centrality} Eigenvalues and Eigenvector are one of the most popular analytical tool to understand behaviour of a square matrix and its linear transformations. Bonacich's~\cite{Bonacich:1972} proposed that the eigenvector corresponding to the largest eigenvalue of a network's adjacency matrix may also be considered for ranking nodes. This measure assigns importance of a node proportional to the sum of the importance of neighbors of that node. The eigenvector centrality of a node $i$ is defined as, $$EC(i)=\sum_{j\in V\setminus\{i\}}( a_{ij} \cdot EC(j)),$$ where recall that $a_{ij}$ denotes the adjacency relationship between node $i$ and $j$. Eigenvector centrality resolves the local view based limitation of degree centrality. This measures assumes that if a person's friends are powerful in the network then that person will also be powerful. Nodes with higher eigenvector centrality scores denote that such nodes have connection to other powerful nodes in networks. A person with lower eigenvector centrality in a social network denotes that the friends of that person are not important and powerful. The major limitation of this measure is that it does not work well in directed acyclic networks. This measure gave basis to define one of the most popular and extensively used ranking measure PageRank which is used in Google to rank pages before giving search results. Few other centrality measures have been developed on similar principle to eigenvector centrality which have been proven to be extremely usable for network analysis. \\

Several studies have analysed and compared the above mentioned traditional centrality measures~\cite{Borgatti:2005,Borgatti:2006,Landherr:2010,Das:2018}. It has been observed that although the top central nodes as per these measures may differ on various networks, but the ranking of all the nodes by these measures are positively correlated~\cite{Lee:2006,Valente}. Ranking due to degree centrality has been found to be highly correlated with the ranking due to betweenness and eigenvector centrality measures. In the next section, we summarize few other centrality measures that have been extensively used to analyse social and complex networks.\\
\subsection{Other Popular Centrality Measures}
This section mentions few other popular centrality measures other than the traditional ones which are used to analyse social and complex networks.

\subsubsection{Katz Centrality:} This measure can be used to estimate the influence of a person in a social network. According to this measure, the importance of a node is not a mapping based on the number of neighbours or the shortest path lengths. It considers the number of walks between node pairs to assign importance~\cite{Katz:1953}. Mathematically, it is defined as {\footnotesize $$KC(i)=\sum_{k=1}^{\infty}\sum_{j\in V} \alpha^k (A^k)_{ij} ,$$} where $(A^k)_{ij}$ denotes the total number of walks of length $k$ from node $i$ to node $j$ and $\alpha$ represents an attenuation factor that helps damping the effect of longer walks while computing the importance. The value of the attenuation factor is chosen such that $0 \le \alpha\le \frac{1}{\lvert \lambda \rvert}$ where $\lambda$ denotes the principle eigenvalue of adjacency matrix $A$. Few variations and generalizations of this measure are given in \cite{Hubbell:1965,Bonacich:1987,Bonacich:2001}.

 \subsubsection{PageRank Centrality:}  This centrality measure~\cite{Brin:1998,Page:1999} was introduced for the directed web-page network to rank web-pages for efficient searching. Google search engine came into light after this measure. Katz centrality faced an issue that if a high central node points to many other nodes, then all of those nodes also attain high centrality score. PageRank resolves this issue by diluting the contribution of the neighbouring nodes using their out-degrees. The PageRank centrality of a node $i$ is defined as: {\footnotesize$$PRC(i) = \alpha \sum_{j\in V} \frac{a_{ij}}{D_j} \, PRC(j) + \beta,$$} where $\alpha$ and $\beta$ are two constant quantities and $D_j$ denotes the number of links outgoing from node $j$. Whenever there are no outgoing links from a node $j$, $D_j$ is considered 1. $\alpha$ and $\beta$, similar as considered in \cite{Hubbell:1965,Bonacich:1987,Bonacich:2001}, are the factors for consideration of dependency on the network topology and exogenous component respectively. 
 
 \subsubsection{Decay Centrality:} This centrality is similar to closeness and harmonic centrality and is also based on the shortest path lengths to all the nodes in a network. Harmonic centrality computes importance as the sum of inverse of distances while this measure assigns importance proportional to the sum of an exponentially decreasing function over distance~\cite{Jackson}. It is defined as
{\footnotesize $$DKC(i)=\sum\limits_{i\in V\setminus\{j\}} \delta^{d_{ij}}$$} 
where $\delta$ is a decay parameter such that $0< \delta < 1$. This measure can be used in applications where in the place of harmonic or closeness centrality, the geodesic distances have to be penalized exponentially.
 
 \subsubsection{Social Centrality:} Recently, Saxena et al.~\cite{Saxena1:2018} proposed a new centrality measure specific to social networks and called it \textit{social centrality}. This measure assigns importance to a node proportional to its socializing capability to gain access of resources available on other nodes and its inter-community/intra-community ties which represent its bonding potential within its community and bridging potential to other communities. The centrality score of a node is computed by aggregating its sociability index with its bridging and bonding potential. A high central node as per this measure can easily manage access to resources available within the system due to its hierarchy and position within and across communities while a low central node may struggle for the resources. 
 
 \subsubsection{Other Centrality Measures:} Few other popular measures for analyzing social networks are mentioned next. \textit{Information Centrality}~\cite{Stephenson:1989} is based on the all possible paths between pairs of points and the information contained on these paths. Hage and Harary\cite{Hage95} introduced \textit{eccentricity} as a centrality measure which gives larger importance to the node whose maximum geodesic distance to other nodes in the network is smaller. Brandes and Fleischer\cite{Brandes1:2005} defined variations of closeness and betweenness centrality called \textit{current-flow closeness} and \textit{current-flow betweenness} which assumes that spread of information is like electricity, therefore, not only the shortest paths, but all possible paths should be considered. They showed that current-flow closeness measure is same as information centrality~\cite{Stephenson:1989}. \textit{Diffusion centrality} is a measure to evaluate the influence of actors in a social network for spreading information~\cite{Banerjee:2013}. It is a generalization of degree, eigenvector and katz centrality. \textit{Coverage centrality}~\cite{Yoshida:2014} is similar to betweenness centrality and it assigns importance to a node proportional to the number of pairs of nodes between which at least one shortest path passes through that node.

\section {Directions of Research} In this section, we discuss various directions of research related to centrality measures in social and complex networks. We start with approaches for exact computation of traditional measures. The computation of few kinds of centrality scores has been realized to be expensive in terms of time over large networks. Several studies have been conducted that focused on fast estimation of those types of centrality scores to tackle the issue. We summarize few of such literature on estimation of tradition centrality measures. Next, we focus on the problem of computing and keeping centrality measures up to date in dynamic networks. These type of networks evolve over time. Algorithms for such kind of networks are called dynamic algorithms and we brief few related literature on centrality measures. Few of the recent studies on estimation algorithms over dynamic networks are mentioned next. Afterwards, parallel and distributed algorithms for speeding up and scaling computation of traditional measures are mentioned. Although, computing top-$k$ central nodes as per a centrality measure is a widely studied problem but recently researchers started designing fast algorithms for ordering/ranking a set of arbitrary nodes in a large network based on some centrality measure. We note down few studies on both types of problems. Further, few generalizations of centrality measures considering weights either on the edges or on the nodes or on both have been discussed. Some applications require computing cumulative centrality scores of a set of nodes than computing individual scores. These measures are know as group centrality and few related studies are briefed next. Hybridization of centrality measures to analyze social and complex networks is another direction. A graph-editing based problem on improvement or maximization of centrality scores is discussed next. Finally, some applications of centrality measures in social networks are summarized.
\subsection{Exact Computation}
In this section, we discuss algorithms that compute exact traditional centrality scores of nodes. The exact algorithm to compute degree centrality is very trivial which requires $O(n)$ time to compute the degree of a node and $O(m)$ time to compute  the degree of all nodes. Recall that $n$ denotes the number of nodes (order of a network) and $m$ denotes the number of links (size of a network). A simple exact algorithm to compute closeness centrality score of a node is based Dijkstra' single source shortest path (SSSP) computation algorithm \cite{Dijkstra:1959} which takes $O(m+n\log n)$ and $O(m)$ time in weighted and unweighted networks respectively. Closeness scores of all nodes can be computed using either SSSP computation from all nodes requiring $O(mn+n^2\log n)$ and $O(mn)$ time in weighted and unweighted networks respectively or all pair shortest path (APSP) computation using Floyd-Warshall's algorithm~\cite{Floyd:1962,Warshall:1962} which takes O($n^3$)time. Sariy{\"u}ce et al.~\cite{Sariyuce:2017} proposed a framework to compute closeness centrality faster than the trivial approach. Their proposed framework modifies a network by compressing and splitting it into small sub-networks in which centrality scores can be computed independently. Their proposed algorithm empirically outperformed competitive algorithms by several folds.\\

Kintali~\cite{Kintali:2008} conjectured that exact betweenness score computation of a node is as time consuming as computing betweenness scores of all nodes. Similar to closeness centrality, all the algorithms to compute betweenness scores are either based on SSSP computation from all nodes or APSP computation. A modified version of the Floyd-Warshall's APSP computation algorithm ~\cite{Floyd:1962,Warshall:1962} is the most trivial algorithm to compute exact betweenness scores for one as well as all nodes. As stated above, this approach takes $O(n^3)$ time. Computation of betweenness scores takes $O(n^3)$ even when SSSP computation from all nodes are used. It is due to the reason that even when the number of shortest path between all pair of nodes are given, computation of betweenness formulation still takes $O(n^3)$ time. Brandes~\cite{Brandes:2001} reformulated the definition of betweenness centrality in terms of summing up \textit{dependency} value. Dependency of a node $i$ on node $j$ denotes the contribution of the shortest paths originating at node $i$ in the betweenness score of node $j$. His algorithm was based on a modification to  Dijkstra's\cite{Dijkstra:1959} algorithm. Due to the new formulation, it started computing exact betweenness score in the same asymptotic time whatever was required for running Dijkstra's\cite{Dijkstra:1959} algorithm from all nodes. Although, several faster algorithms by Baglioni et al.~\cite{Baglioni:2014}, Puzis et al.~\cite{Puzis:2015}, Sariy{\"u}ce et al.~\cite{Sariyuce:2013}, Erdos et al.~\cite{Erdos:2014}, Chehreghani et al.~\cite{Chehreghani:2018}, Bentert et al.~\cite{Bentert:2018}, and Daniel et al.~\cite{Daniel:2019} have been proposed that attempted to reduce the time to compute betweenness score empirically or theoretically on some special type of networks, but so for no algorithm guarantees to perform asymptotically better than the algorithm by Brandes~\cite{Brandes:2001}.\\ 

Eigenvector centrality scores can be computed using the power method~\cite{Wilkinson:1965}. The power method starts with a vector whose euclidean norm is 1 as an initial approximation of the eigenvector corresponding to the largest eigenvalue. Each iteration of this method takes the resulting vector from previous iteration as input and multiplies it with the adjacency matrix of the network under consideration to improve the approximation. The convergence of this method is certain if the adjacency matrix has a dominant eigenvalue. The time for convergence depends on the ratio between the absolute values of the dominant and the second dominant eigenvalues. The same method is also used to compute PageRank and some other variants of eigenvector centrality. A basic foundation for algorithms to compute traditional centrality scores is given in Chapter~4 in the book by Brandes and Erlebach~\cite{Brandes:2005}

\subsection{Estimation \label{estimation}} Due to the large size of social networks, even the best algorithms for computing exact centrality scores might be time consuming. To overcome this limitation, researchers developed several estimation (approximation) algorithms that take relatively lesser amount of time than exact algorithms and compute approximate values of centrality scores. Most of the estimation approaches are sampling based. In the sampling technique, in place of conducting computation based on every member from a set of entities, a subset of entities are chosen and then estimated values are computed based on that subset. Sampling may be uniform or nonuniform. In this section, we briefly discuss few such studies. The exact computation of the degree centrality of a node as well as all the nodes is very efficient, therefore, there does not seems any requirement of an estimation algorithm. Though when one wants to know a node's rank using only the local information, a need of a rank estimation algorithm arises even for degree centrality. Details about rank estimation algorithms are given in Section~\ref{ordering}.\\

Eppstein and Wang~\cite{Eppstein:2004} proposed a node sampling based algorithm to estimate closeness centrality scores of all the nodes and gave theoretical bounds on the error in estimating scores. The idea was to sample a few nodes from the set of all nodes and consider the single source shortest path computation (SSSP) from only the chosen nodes(also called pivot) for centrality computation. Ohara et al.~\cite{Ohara:2014} proposed a similar algorithm to estimate closeness centrality scores as by Eppstein and Wang~\cite{Eppstein:2004}, but they gave a different theoretical analysis than \cite{Eppstein:2004}. Rattigan et al.~\cite{Rattigan:2006} proposed to create network structure index for efficient estimation of closeness centrality and betweenness centrality. Cohen et al.~\cite{Cohen:2014} gave a scalable algorithm for estimating closeness centrality scores on undirected as well as directed graphs. A group testing based algorithm for identifying top closeness central nodes was given by Ufimtsev and Bhowmick~\cite{Ufimtsev:2014}. Murai~\cite{Murai:2017} gave pivot guided estimation based algorithm to estimate closeness centrality scores in undirected networks and strongly connected directed networks. His algorithm outperformed the estimation algorithms in \cite{Eppstein:2004,Cohen:2014} theoretically as well as empirically .\\

Computing one node's betweenness centrality has been conjectured to be as time consuming as computing betweenness scores of all the nodes. There are two classes of estimation algorithms for betweenness centrality. The first one focuses on estimating the scores of all the nodes together while the other one just estimates betweenness score of a particular node. Brandes and Pich~\cite{Brandes:2007} used a similar idea as used by Eppstein and Wang ~\cite{Eppstein:2004}, for estimating betweenness centrality measure. An adaptive node sampling based algorithm was proposed by Bader et al.~\cite{Bader:2007} to estimate a node's betweenness score. A theoretical bound on the error was also provided. Geisberger et al.~\cite{Geisberger:2008} proposed a generalization of the algorithm coined by Brandes and Pich\cite{Brandes:2007} which achieved better results. Most of the studies discussed use randomization algorithms, but Gkorou et al.~\cite{Gkorou:2010} and Ercsey-Ravasz et al.~\cite{Ercsey:2012} proposed a deterministic estimation algorithm. They gave an estimation algorithm for computation of the betweenness scores in large networks by considering only the shortest paths of length $k$. A comparative analysis of Gkorou et al.'s algorithm~\cite{Gkorou:2010} with Geisberger et al.'s~\cite{Geisberger:2008} and Brandes and Pich's~\cite{Brandes:2007} algorithm is done in \cite{Gkorou:2011}. Ohara et al.~\cite{Ohara:2014} also studied estimation of betweenness centrality in addition to closeness centrality and did bound analysis for error in estimation. Riondato and Kornaropoulos~\cite{Riondato:2014} proposed two randomized algorithms based on the sampling of shortest paths to estimate betweenness scores. Chehreghani~\cite{Chehreghani:2014} used non-uniform node sampling to estimate a nodes' betweenness score. Agarwal et al.~\cite{Agarwal:2015} analysed random graphs and proposed another non-uniform node sampling based estimation algorithm which performed better than \cite{Chehreghani:2014} and other competitive algorithm to estimate a node's betweenness centrality score. Their estimation algorithm was further applied to solve betweenness-ordering problem\cite{Singh:2018}. Due to popularity and wide applicability, most of the estimation algorithm for centrality measures are for betweenness measure. Several recent studies approximately compute betweenness scores\cite{Ostrowski:2015,Furno:2017,Furno:2018,Borassi:2019,Haghir:2019,Chehreghani:2019}. A review of approximation algorithms for computing betweenness centrality has been done by Matta et al.~ \cite{Matta:2019}.

Wink et al.~\cite{Wink:2012} presented an algorithm to estimate voxel-wise eigenvector centrality scores in fMRI data. Kumar et al.~\cite{Kumar:2015} gave an estimation algorithm for eigenvector centrality and PageRank based on neural networks. Charalambous et al.~\cite{Charalambous:2016} proposed a distributed approach to efficiently estimate eigenvector centrality of nodes in directed networks. Ruggeri and De Bacco~\cite{Ruggeri:2019} gave an algorithm on incomplete graphs to estimate eigenvector centrality scores. Their estimation algorithm is based on a sampling idea  derived from spectral approximation theory. Mitliagkas et al.~\cite{Mitliagkas:2015} proposed a fast approximation algorithm to estimate PageRank.

\subsection{Updating Centrality Scores} Real-world Networks are large in size and dynamic in nature. Therefore, to maintain updated centrality scores of nodes, applying exact algorithms after every or even few number of updates in batches can be impractical when the exact algorithms are time consuming on large networks. There can be a significant difference in the ranking of vertices before and after an update~\cite{Yan:2010}. Updates can be insertion/deletion of edges or nodes or increase/decrease in edge weights. Algorithms designed to update values of some attributes on nodes/edges or some other network properties in case of updates in networks faster than re-computing scores using exact algorithms are called \textit{dynamic algorithms}. Algorithms tackling different nature of updates are categorized as incremental, decremental or fully dynamic algorithm. In this section we briefly mention some dynamic algorithms for traditional centrality measures.\\

Kas et al.~\cite{Kas1:2013} proposed an incremental algorithm to update closeness scores in evolving social networks after addition/removal of links and nodes. Sar{\i}y{\"u}ce et al.~\cite{Sariyuce1:2013} gave an incremental algorithm for computing closeness scores after edge insertion / deletions. Yen et al.~\cite{Yen:2013} also proposed a dynamic algorithm to update closeness centrality scores after edge insertion/deletion. The basic idea used in their algorithms is to efficiently identify nodes whose closeness centrality will change after a link update. Wei and Carley~\cite{Wei:2014} proposed an online algorithm framework to update closeness and betweenness scores after link updates. Khopkar et al.~\cite{Khopkar:2014} proposed an incremental algorithm for all pair shortest paths and used the idea to develop incremental algorithm for closeness and betweenness centrality. Sar{\i}y{\"u}ce et al.~\cite{Sariyuce:2015} also gave an incremental algorithm to compute closeness centrality scores in dynamic networks relying on a distributed memory framework. Santos et al.~\cite{Santos:2016} proposed a scalable algorithm for updating closeness centrality scores after deletion of edges. A dynamic algorithm to compute the closeness centrality of a node in social networks that are evolving with time, is given by Ni et al.~\cite{Ni:2019}. Most of the algorithms to compute closeness centrality are based on network topology and structures while their idea relies on temporal network features. Shao et al.~\cite{Shao:2020} recently gave a dynamic algorithm to compute closeness. They proposed to calculate the  exact closeness centrality scores by using bi-connected blocks and articulation vertices. Their approaches is to detect all shortest paths that are affected and then update the centrality value based on articulation vertices.\\

Vignesh et al.~\cite{Vignesh:2011} considered the problem of updating betweenness scores after addition or deletion of nodes. Lee et al.~\cite{Lee:2012} proposed a dynamic algorithm for betweenness centrality to tackle link updates (addition/deletion). Their approach was based on the observation that whenever an update happens within a bi-connected component, the re-computation of centrality scores are required only for the nodes in that bi-connected block. Betweenness scores of nodes outside that block can be updated very efficiently without a need of re-computation. For node updates, they suggested to use their proposed algorithm for every link incident on nodes under consideration. Green et al.~\cite{Green:2012} proposed an incremental algorithm for betweenness centrality in case of a series of link addition over time. The idea used in their algorithm was to maintain and update breadth first tree data structure rooted at every vertex. Kas et al.~\cite{Kas:2013} also gave an incremental algorithm to tackle updates for nodes and edges in the form of change in edge weights. Their idea was based on Ramalingam and Reps'~\cite{Ramalingam:1991} incremental approach  for updating all pair shortest paths (APSPs). Further they extended their incremental approach for a variant of betweenness centrality where the shortest paths of at most $k$ lengths are considered for computation of betweenness centrality scores \cite{kas:2014}. An incremental algorithm similar to the one in \cite{Green:2012} was given by Nasre et al.~\cite{Nasre1:2014}. Their algorithm tackled node as well as edge updates and used breadth firsts search based directed cyclic graphs (BFS DAGs) as the data structure in place of BFS trees. Later, Nasre et al.~\cite{Nasre:2014} proposed a decremental approach for updating all pair all shortest paths (APASPs) extending the approach by Demetrescu and Italiano~\cite{Demetrescu:2004} for APSPs which founded basis for a decremental approach to update betweenness scores. Kourtellis et al.~\cite{Kourtellis:2014} proposed a scalable online algorithm to update betweenness scores of nodes and edges in case of edge addition/deletion. Pontecorvi and Ramachandran~\cite{Pontecorvi:2014} gave a fully dynamic algorithm for updating APASPs and extended the algorithm to develop a fully dynamic approach to update betweenness scores~\cite{Pontecorvi:2015,Pontecorvi1:2015}. A dynamic algorithm to update betweenness scores after node addition and deletion, similar to Lee et al.'s~\cite{Lee:2012} approach, was given in \cite{Goel:2013,Singh:2015}. Hayashi et al. \cite{Hayashi:2015}, Bergamini et al.~\cite{Bergamini:2017} and  Tsalouchidou et al.~\cite{Tsalouchidou:2019} also gave dynamic algorithms to update betweenness scores.\\

More work has been done to update PageRank Centrality and Katz Centrality in dynamic networks than traditional Eigenvector Centrality. Bahmani et al.~\cite{Bahmani:2010}, Rossi and Gleich~\cite{Rossi:2012}, Rozenshtein et al.~\cite{Rozenshtein:2016}, and recently Zhan et al.~\cite{zhan:2019} gave algorithms to update PageRank in dynamic networks. Nathan and Bader~\cite{Nathan1:2017} gave an algorithm to update Katz centrality scores in a dynamic network. There has been studies to update various other centrality scores in dynamic networks. For example, Sarmento\cite{Sarmento:2017} gave an incremental algorithm to update laplacian centrality measures. Recent studies on updating centrality scores in dynamic networks show that the direction is still open for more efficient algorithm for various centrality measures.\\
 
 Although dynamic graphs and algorithm on dynamic graphs have been studied extensively in the last few decades, in the last decade, it has been explored in several studies under a new name called \textit{temporal networks}~\cite{Holme:2012,Lambiotte:2016} or time-dependent graphs~\cite{Wang:2019}. On such types of networks, the question of identifying important nodes and edges are done with the help of centrality measures for Temporal networks~\cite{Kim:2012}. For example, closeness centrality~\cite{Pan:2011}, betweenness centrality~\cite{Tsalouchidou:2019}, eigen-vector centrality~\cite{Taylor:2017}, random-walk centrality~\cite{Rocha:2014}, pagerank centrality~\cite{Lv:2019}, etc. have been studied over temporal networks. 

\subsection{Approximation Algorithms for Dynamic Graphs}
Several literature on centrality measures studied either dynamic algorithms or approximation algorithms for computation of centrality scores in the last two decades. Recently, the problem of updating estimated centrality scores in dynamic networks came into light. Bergamini et al.~\cite{Bergamini:2014}, Bergamini and Meyerhenke ~\cite{Bergamini:2015},Riondato et al.~\cite{Riondato:2018} and Chehreghani et al.~\cite{Chehreghani1:2018} proposed algorithms for updating approximated betweenness centrality scores in dynamic networks. Zhang et al.~\cite{Zhang:2016} gave such kind of approach for personalized pagerank centrality while Nathan and Bader~\cite{Nathan:2017} gave for personalized katz centrality measure.

\subsection{Parallel and Distributed Computation}
Real-world networks are very large in size. The computation of  centrality scores for closeness, betweenness and similar measures require asymptotically quadratic or cubic time in the order of networks. These kind of computations are time consuming when implemented sequentially. Similar order of time is required to keep the scores up to date when network topology changes over time. Parallel computing has been proven as one of the best methods to reduce time for computation whenever algorithms support parallelism by utilizing super-computing resources. Distributed computing is a popular tool to perform large-scale computation. Distributed algorithms for centrality measures aim to compute centrality scores at each node using information attained by those nodes based on the interactions with their neighbors. Due to this reason. Distributed algorithms also face a challenge to exactly compute those centrality measures that require information of the whole network.\\

Several literature in the last two decades study parallel and distributed computation of centrality measures over static as well as dynamic networks. Bader et al~\cite{Bader:2006} studied parallel algorithms to compute degree, closeness and betweenness centrality. A recent study on parallel computation of these measures is \cite{Garcia:2019}. \cite{Santos:2006,Santos:2016} gave algorithms for computation of closeness centrality in a parallel setting. Shukla et al.~\cite{Shukla:2020} gave parallel algorithms for closeness and betweenness centrality in dynamic networks. \cite{Wang:2014,Wang1:2015} have given distributed algorithms for tree and general networks. You et al. \cite{You:2016} have studied distributed computation for the degree, closeness, betweenness, and PageRank centrality. Most of literature related to parallel and distributed compution of centrality measures are for betweenness centrality. It is due to the factor that even the computation of betweenness centrality of a node is time consuming and the scalability of the computation is challenging. Following are literature on computation of exact and approximate betweenness scores in parallel~\cite{Kintali:2008,Madduri:2009,Edmonds:2010,Sariyuce1:2013,Prountzos:2013,Mclaughlin:2014,Bernaschi:2016,Solomonik:2017,Castiello:2018,Vella:2018,Grinten:2019,Van1:2019,Daniel:2019,Grinten:2020} and distributed \cite{Wang:2013,Wang:2015,Hoang:2019,Crescenzi:2020} frameworks.

\subsection{Centrality Ordering and Ranking\label{ordering}}

Most of the algorithms for centrality measures compute or estimate scores to rank nodes. Some applications may demand ranking of top $k$ nodes with high centrality scores while others may want to rank a set of arbitrarily picked nodes. The first problem is called \textit{top-$k$ central node} computation while the later one is known as \textit{centrality-ordering} problem~\cite{Singh:2018}.
A solution to the above problems may output exact or estimated ranks. Several studies on ranking all nodes, estimating a node's rank, finding top-$k$ central nodes or ordering $k$ arbitrarily picked nodes based on various centrality measure have been conducted. \\

Bian et al.~\cite{Bian:2019} recently conducted a survey on identification of the top $k$ nodes based on degree centrality, closeness centrality, and influence for diffusion. Studies on ranking of the top $k$ central node based on closeness centrality\cite{Okamoto:2008,Bisenius:2018}, betweenness centrality\cite{Riondato:2014,Mahyar:2018,nakajima:2020}, and 
katz centrality~\cite{Zhan:2017} may be referred. Kumar et al.~\cite{Kumar:2015} gave rank estimation algorithm on the basis of eigenvector centrality and PageRank based on neural networks. Computation of degree centrality of a node is very efficient but identifying rank of a node based on degree centrality requires larger computation. Saxena et al.\cite{Saxena:2018} proposed methods to estimate a node's degree rank. Computing Closeness centrality of a node takes relatively a lot smaller time than closeness rank of that node. Saxena et al.\cite{Saxena:2019} gave a heuristic to estimate a node's closeness rank. Kumar et al.~\cite{Kumar:2015} gave a neural networks based rank estimation algorithm on the basis of eigenvector centrality and PageRank.Singh et al.~\cite{Singh:2018} introduced centrality-ordering problem and gave an efficient algorithm to estimate betweenness-ordering. They motivated for an open direction related to study of the ordering problem on other centrality measures.

\subsection{Weighted Centrality Measures}

The common practice of defining centrality measures is to first introduced it for unweighted network, i.e. every actor or entity represented by nodes are assumed to have same features and relationships between actors are also assumed to be uniform. These measures are called unweighted centrality measures. Definitions of the traditional centrality measures given in Section~\ref{trad} are for unweighted networks. In some of the networks, weights on the edges are given and to better analyze the network, it becomes essential to use the weights. The definition of centrality measures that considers weights on the edges while computing the scores are called edge-weighted centrality measures. Although, the weights on the edges are taken into account for analysis, yet the weights on nodes are still assumed to be uniform. Most of the weighted version of the centrality measures are defined only considering the edge weights.~\cite{Newman:2001,Lee:2006,Opsahl:2010,Qi:2012,Wei:2013}. 
The edge-weighted degree centrality has been used in several applications in biological network to identify crucial nodes~\cite{Li:2010,Tang:2013,Candeloro:2016}. \\

Similarly, some studies defined node-weighted centrality measures by considering weights on the nodes and uniform  weights on the edges\cite{Abbasi:2013,Abbasi1:2013,Wiedermann:2013,Akanmu:2014,Singh:2019}. These studies suggested to combine the edge-weighted version of the definition of the centrality measures to get fully-weighted centrality measures that can analyze networks while considering weights on the edges as well as on the nodes. The assumption of uniform weights on the nodes and the edges is to simplify the analysis of networks. It is highly unlikely that all the actors in a network possess same characteristics and features. We are surrounded by fully weighted networks but edge weights are easily available in comparison to node weights. Due to privacy and security concerns, actor do not share details about their personal information which makes it difficult and complicated to map characteristics / attributes / features of actors in the form of node weights. Relationship data is relatively easily available and, therefore, it becomes relatively easy to consider edge-weights. In other type of complex networks, similar constraints exist. Singh et al.~\cite{Singh:2019} proposed a way to overcome the difficulty in figuring out weights on the nodes. They suggested to apply appropriate measures to generate weights on nodes and then apply node-weighted or fully-weighted centrality measures.

\subsection{Group Centrality Measures} 
Everett and Borgatti~\cite{Everett:1999} extended the idea of centrality measures for individual nodes/edges to compute collective centrality scores of a group of nodes/edges. These type of centrality measures identify a set/group/class of nodes or edges which collectively dominate other sets/groups/classes on the basis of a quantitative notion of importance. An application oriented study of the this variant of degree, closeness, and betweenness centrality has been conducted by Ni et al.~\cite{Ni:2011}. Zhao et al.~\cite{Zhao:2014} gave an efficient algorithm to compute group closeness centrality for disk-resident networks. Chen et al.~\cite{Chen:2016} shown that the problem of finding a group of $k$ nodes whose collective closeness centrality is maximum, is a NP-hard problem to solve. Group betweenness centrality has been used in multiple applications~\cite{Dolev:2009,Halappanavar:2012} and to compute or estimate group betweenness centrality, several algorithms have been proposed~\cite{Puzis:2007,Brandes:2008,Kolaczyk:2009,Chehreghani1:2018}.
\subsection{Hybrid Centrality Measures}
Individual centrality measures might not appear as a fruitful tool for analysing some complex systems which has a mixed notion of importance. For networks based on such systems, hybrid centrality measures are used. Hybrid centrality measures are defined by combining more than one measure to produce better rank than individual ranks by each measure in the combination. In this section we briefly mention few literature on hybridizing centrality measures. Few of these can be used in social network analysis as well.\\

In a recent study by Singh et al.~\cite{Singh:2019}, a new way of centrality hybridization based on the formulation of node-weighted centrality measures was given. Singh et al.~\cite{Singh:2019} proposed to generate weights on nodes based on a centrality measure, and then use the generated weights while computing node-weighted version of another centrality measure. They also applied these measures for two applications. One of the demonstrated applications of such hybridization was to find influential spreaders in a complex contagion scenario. Abbasi and Hossain~\cite{Abbasi:2013} proposed a new set of hybrid centrality measures by hybridizing degree, closeness, and betweenness centrality measures within the framework of degree centrality. They applied their hybrid measures on a real-world co-authorship network and noted that the hybrid measures performed differently than the traditional centrality measures and further noticed that the newly proposed measures were significantly correlated to authors' performance. Abbasi~\cite{Abbasi1:2013} proposed hybrid measures on weighted collaboration networks based on h-index, a-index, g-index. These indices (h-index, a-index, g-index) are considered as traditional collaborative performance measures to rank authors. The proposed measure gave results highly correlated with ranking based on citation-count and publication-count. The two quantities, citation-count and publication-count, are widely used and well established performance measures for scholars. All of the three studies mentioned above proposed hybrid measures that has application for analyzing social networks of different nature. \\

Linear combination is a popular strategy for combining values across several disciplines. Qiu et al.~\cite{Qiu:2014} has defined a hybridization based on this principle to mix cohesion centrality and degree centrality. This hybridization was further used by Li-Qing et al.~\cite{Li-Qing:2014} for detecting community structures. A hybridization of closeness centrality and betweenness centrality was proposed by Zhang et al.~\cite{Zhang:2014} rank nodes in satellite communication networks. In another study, a hybridization of degree centrality, a variation of traditional closeness centrality for disconnected networks, and betweenness centrality measures was proposed by Buechel and Buskens~\cite{Buechel:2013}. A hybrid page-ranking approach based on the traditional centrality measures was proposed by Qiao et al.~\cite{Qiao:2010}. Lee and Djauhari \cite{Lee:2012} also had proposed a linear combination based hybridization  oftraditional centrality measures which was applied to identify highly significant and influential stocks. In an early study by Wang et al. \cite{Wang:2008}, a hybridization of degree centrality, betweenness centrality, and degree of neighbors was proposed.

\subsection{Centrality Improvement and Maximization}
 A graph editing problem related to centrality measures is to improve or maximize centrality score of a node by adding links. Several literature in the last two decades study centrality improvement or maximization problems. Avrachenkov and Litvak~\cite{Avrachenkov:2006} studied the change in pagerank scores due to link addition. Later page-rank maximization problem using addition of new outgoing links\cite{Kerchove:2008} or new incoming links \cite{Olsen:2010} was considered. Maximization of eccentricity centrality~\cite{Demaine:2010,Perumal:2013} was studied soon after. Further, the centrality improvement and maximization problem was considered for other centrality measures: Closeness and Harmonic centrality`\cite{Crescenzi:2016}, Betweenness centrality~\cite{Angelo:2016,Bergamini:2018},Information Centrality~\cite{Shan:2018}, and Coverage centrality~\cite{Medya:2018,Angelo:2019}. This nature of graph editing problem has also been explored for maximization of group centrality measures~\cite{Medya:2018,Angriman:2020}.

\subsection{Application} Centrality measures have been widely used to analyse social and complex networks. In this section, we brief few application on social networks. Girvan and Newman~\cite{Girvan:2002} have suggested to use edge betweenness centrality to detect community structures in social and complex networks. Yan and Ding~\cite{Yan:2009} have applied degree, closeness, betweenness, and PageRank centrality in a co-authorship network for impact analysis. Ghosh and Lerman~\cite{Ghosh:2010} have analysed that a variation of katz centrality turns out to be a good predictor of influence in online social networks. Ilyas and Radha~\cite{Ilyas:2011} have used centrality measures to identyfy influential nodes in a online friendship network from Orkut and a gaming network from Facebook. Mehrotra et al.~\cite{Mehrotra:2016} have proposed to use centrality measures for detection of fake followers on Twitter. Riquelme and Gonz{\'a}lez-Cantergiani~\cite{Riquelme:2016} have conducted a survey on various measures including centrality to evaluate user's influence on Twitter.\\

Few of the recent applications of centrality measures are summarized next. Eigenvector centrality can be used to analyse fMRI data of the human brain to identify connectivity pattern~\cite{Lohmann:2010}. In a  study by Zinoviev~\cite{Zinoviev:2020}, a social network is formed of Russian Kompromat has been analyzed using the traditional centrality measures which identified Vladimir Putin as the top central kompromat figure. Kim et al.~\cite{Kim:2020} used normalized closeness and betweenness centrality measures on a word network derived from users’ posts on Reddit to analyze the perspective of public towards renewable energy and identifying frequent issues related to renewable energy. Nurrokhman et al.~\cite{Nurrokhman:2020} has recently used degree, closeness, and  betweenness centrality to  analyze  the collaboration within students for sharing knowledge.  Stelzhammer in his thesis \cite{Stelzhammer:2020} attempts to improve detection of influential users in a recommender system using centrality measures. Trach and Bushuyev~\cite{Trach:2020} have used degree, betweenness, eigenvector, and pagerank centrality measures to analyse a social network between project participants for construction of a residential building located in Ukraine. Yuan~\cite{Yuan:2020} have used degree centrality and structural holes to analyze and forecast tourist arrivals in a tourism social network. Neuberger~\cite{Neuberger:2020} has analyzed relationship between the actors and directors from Soviet film industry. Nagdive et al.~\cite{Nagdive:2020} have used centrality measures to identify key organizations, places and persons in a terrorist network.

\subsection{Defining New Centrality Measures} The above sections brief about various research directions for computing and applying centrality measures for analysing social and complex networks. The last direction that existed since the beginning of the study on centrality measures is to define a new measure when other measures doesn't seem useful enough. This had led us to a point today when there is an abundance of centrality measures. A web-page (\textit{http://schochastics.net/sna/periodic.html}) contains a list of several centrality measures in an interesting representation. It remains open to define new centrality measures that perform better than the existing measures and provide more insightful analysis of social and complex systems. 
\section {Conclusion}
 Centrality measures have been a popular tool to mine social network data. In this chapter, we have reviewed various directions of research related to computing centrality measures and applying these in identification of key actors in social as well as complex networks. Most of the research directions are still evolving and comprise open problems to improve existing approaches, design new algorithms that outperform previous ones, and solve new problems related to computation of various centrality measures. 

\bibliographystyle{plain}
\bibliography{Centrality}
\end{document}